\documentclass[twocolumn,prb]{revtex4}
\usepackage{times, mathptmx, graphicx, amsmath, amssymb, bm, units}

\DeclareMathOperator{\sech}{sech}
\DeclareMathOperator{\sgn}{sgn}

\renewcommand{\Im}{\text{Im}}

\newcommand{\DT}[1][ ]{\mathcal{D}_{T#1}}
\newcommand{\DL}[1][ ]{\mathcal{D}_{L#1}}
\newcommand{\T}[1][ ]{\mathcal{T}_{#1}}

\newcommand{\Rc}[1][ ]{\mathcal{R}_{#1}}

\newcommand{\jS}[1][ ]{j_{S#1}}

\newcommand{\abs}[1]{\left\lvert{#1}\right\rvert}

\newcommand{\dd}[1]{\mathrm{d}#1\,}

\newcommand{\authors}[1]{{\slshape(AUTHORS: #1)}}

\renewcommand{\authors}[1]{}

\newcommand{\mat}[1]{\begin{pmatrix}#1\end{pmatrix}}
\newcommand{\uvec}[1]{\hat{\mathbf{#1}}}

\begin{document}

\title{Thermoelectric effects in superconducting proximity structures}
\date{\today}

\author{Pauli Virtanen and Tero T. Heikkil\"a}

\affiliation{Low Temperature Laboratory, Helsinki University of
  Technology, P.O. Box 3500 FIN-02015 TKK, Finland}

\begin{abstract}

Attaching a superconductor in good contact with a normal metal makes
rise to a proximity effect where the superconducting correlations
leak into the normal metal. An additional contact close to the first
one makes it possible to carry a supercurrent through the metal.
Forcing this supercurrent flow along with an additional
quasiparticle current from one or many normal-metal reservoirs makes
rise to many interesting effects. The supercurrent can be used to
tune the local energy distribution function of the electrons. This
mechanism also leads to finite thermoelectric effects even in the
presence of electron-hole symmetry. Here we review these effects and
discuss to which extent the existing observations of thermoelectric
effects in metallic samples can be explained through the use of the
dirty-limit quasiclassical theory.

\end{abstract}

\maketitle



\section{Introduction}
\label{sec:intro}
\authors{Tero}

Applying a bias voltage or a temperature gradient across a conductor
makes rise to charge and energy currents. The linear response
between the biases and currents is described via the thermoelectric
matrix, whose diagonal parts are the charge and thermal
conductances, and the off-diagonal parts are often referred to as
the thermoelectric coefficients. In typical metals, the latter arise
due to the asymmetry between positive- and negative-energy
excitations with respect to the Fermi energy, i.e., electrons and
holes. Such asymmetry in metals is very small, making the typical
thermoelectric effects at sub-Kelvin temperatures hard to measure
accurately.

Placing a superconductor in good contact to a normal-metal conductor
makes rise to finite pair correlations also inside the latter, even
when the pair potential inside it vanishes. This superconducting
proximity effect has an energy-dependent penetration depth; at
typical measurement temperatures of the order of 100 mK it extends
up to the micrometer range. The proximity effect modifies the
thermoelectric response of the normal conductor. Most importantly,
it makes rise to thermoelectric effects which are orders of
magnitude larger than in the absence of superconductivity. The
proximity-induced modifications are discussed in this paper by
employing the quasiclassical theory in the diffusive limit
\cite{usadel70,belzig99,chandrasekhar04}. In this theory, we assume
that all the relevant length scales of the problem exceed especially
the Fermi wavelength (quasiclassical approximation) and the mean
free path (diffusive limit). Examples of such relevant length scales
are the structure size and the superconducting coherence length. A
further property of the quasiclassical theory, especially important
for thermoelectric effects, is that it assumes electron-hole
symmetry. Because of this, in the normal state it predicts vanishing
thermoelectric coefficients.

The proximity modification of the thermoelectric matrix is
conveniently described in Andreev interferometers (see
Fig.~\ref{fig:setup}), where there are two superconducting contacts to
the normal metal. In this structure, the phase difference between the
two contacts affects the proximity modifications, and its presence is
an important requirement for finite thermoelectric effects, at least
within the quasiclassical theory. This type of a dependence of the
electric conductance on the phase has for example been suggested for
use in quantum measurements of flux qubits \cite{petrashov05}.

This paper is organized as follows. In
Sec.~\ref{sec:thermoelectrictransport}, we briefly introduce the
thermoelectric effects and their relations in normal metals and
after that detail the quasiclassical equations for the diffusive
limit. We also discuss briefly effects left out in the present
paper. Section \ref{sec:scspectrum} introduces to the properties of
the spectral supercurrent and two aspects of nonequilibrium
supercurrent: first, we briefly mention the nonequilibrium control
of the supercurrent and then concentrate more on how the energy
distribution function of electrons is controlled with the
supercurrent. The latter is a precursor to the thermoelectric
effects. These are described in Sec.~\ref{sec:thermocoefs}, which
first introduces to the special symmetries of the thermoelectric
matrix, and then details the behavior of its components as a
function of the phase difference and temperature. In
Sec.~\ref{sec:fluxdependence}, we mention the effects relevant when
considering the dependence on the magnetic flux, and finally in
Sec.~\ref{sec:discussion} we conclude and point out the open
questions related with understanding the measurements of the
thermoelectric effects.

\section{Thermoelectric transport in proximity structures}
\label{sec:thermoelectrictransport}
\authors{Tero}

In this paper, we show that the superconducting proximity effect is
able to generate large thermoelectric effects, which can be described
without employing the electron-hole asymmetry. An important factor in
the theory is the presence of supercurrent, which then needs to be
taken into account in the description of the currents. Moreover, to
describe transport between normal metals in the presence of
supercurrent, we need to have multiple terminals connected to the
structure. As the biases mostly deal with the quasiparticle current,
we define the thermoelectric matrix in a multi-terminal structure
according to
\begin{align}\label{eq:linearresponseNS}
  \mat{ I_c^{i} - I_{S,\mathrm{eq}}^{i} \\ I_E^{i} }
  =
  \sum_{j\in\mathrm{terminals}}
  \mat{ L_{11}^{ij} & L_{12}^{ij} \\
        L_{21}^{ij} & L_{22}^{ij} }
  \mat{\Delta V_j \\ \Delta T_j / \bar{T}}
  \,,
\end{align}
where $I_c^i$ and $I_{S,\mathrm{eq}}^{i}$ are the total charge current
and the equilibrium supercurrent flowing to terminal $i$, and
$I_E^{i}$ is the energy current (supercurrent as such carries no
energy current). Moreover, $\Delta V_j$ is the bias voltage and
$\Delta T_j$ is the temperature difference from some average
temperature $\bar{T}$, both present in terminal $j$.

\subsection{Transport in normal-metal structures}
\label{sec:normaltransport}
\authors{Tero}

Thermoelectricity in normal-metal wires can be practically described
especially in the diffusive limit (structure size $L$, elastic mean
free path $\ell_{\rm el}$ and Fermi wavelength $\lambda_F$
satisfying the relation $L \gg \ell_{\rm el} \gg \lambda_F$). In
this limit, the charge and heat currents flowing in the wire are
given by
\begin{subequations}
\begin{align}
I_c &=-eA\int_{-\infty}^\infty dE \tilde{D}(E)\nu(E) \partial_x f(x;E)\\
I_Q &= -A \int_{-\infty}^\infty dE (E-\mu) \tilde{D}(E) \nu(E)
\partial_x f(x;E).
\end{align}
\end{subequations}
Here $\tilde{D}(E)$ is the diffusion constant, $\nu(E)$ is the
density of states, $A$ is the cross-sectional area of the wire and
$x$ is the coordinate parallel to the wire. The heat current can be
simply related to the energy current via $I_Q=I_E-\mu I_c$. For
linear response, $I_Q=I_E$
--- the second term is responsible for Joule heating --- and we can
expand the electron energy distribution function $\partial_x
f=(\partial_T f)
\partial_x T + (\partial_\mu f)\partial_x \mu$, with $f\approx
f_0$, the Fermi function.  Furthermore, assuming some characteristic
length $L$ and taking $\partial_x T=\Delta T/L$ and $\partial_x
\mu=e \Delta V/L$ allows us to relate the results to
Eq.~\eqref{eq:linearresponseNS}.

The energy-dependent changes in the density of states or the
diffusion constant typically take place at large energy scales of
the order of the Fermi energy $E_F$. We can thus expand them as
$\tilde{D}(E) \approx D + c_D (E-E_F)/E_F$ and $\nu(E) \approx \nu_F
+ c_N(E-E_F)/E_F$. To linear order in $c_D$ and $c_N$, we then get
\cite{cutler69}
\begin{subequations}\label{eq:nstateexpressions}
\begin{align}
L_{11}&=G=e^2\nu_F D A/L, \quad &\text{Drude conductance}\\
L_{22}&=L_0 G T^2, \quad &\text{Wiedemann-Franz law}\\
L_{12}&=eL_0 G' T^2, \quad &\text{Mott law}\\
L_{21}&=L_{12}, \quad & \text{Onsager-Kelvin relation}.
\end{align}
\end{subequations}
Here $L_0=\pi^2 k_B^2/(3e^2) \approx 2.45 \times 10^{-8}$
W$\Omega$K$^{-2}$ is the Lorenz number, and the electron-hole
asymmetry is described by the factor $G'=e^2(c_D \nu_F+D c_N)A/(L
E_F)$. These relations show that the thermoelectric effects in
normal metals are of the order of $k_B T/E_F$.

The Onsager-Kelvin relation between the two thermoelectric
coefficients is an example of a more general relation
\cite{onsager31,casimir45,callen48} between different linear-response
coefficients. According to this relation, the elements of the
thermoelectric matrix in Eq.~\eqref{eq:linearresponseNS} should
satisfy
\begin{equation}
L_{\alpha\beta}^{ij}(B)=L_{\beta \alpha}^{ji}(-B)
\end{equation}
under the reversal of the magnetic field $B$. Here $\alpha,\beta \in
\{1,2\}$. This relation results essentially only from the assumption
of time-reversal symmetry. In Sec.~\ref{sec:symmetry}, we show how
this equation can be derived for the energy-dependent response
coefficients within the quasiclassical theory.

The presence of superconductivity modifies the above laws in many
different ways \cite{claughton96}. For example, the Andreev
reflection \cite{andreev64} breaks the Wiedemann-Franz law, and the
Mott law is broken in asymmetric structures \cite{heikkila00}. The
effects related to the superconducting density of states or to
charge imbalance make modifications to the thermoelectric effects at
interfaces \cite{galperin02,schmid79} and for the nonlinear response
\cite{giazotto2006-opportunities}. The main modification at linear
response due to the proximity effect is the appearance of
thermoelectric effects even without electron-hole asymmetry
\cite{seviour00,kogan02,virtanen04,virtanen04b,volkov05,virtanen07}.
The latter effect is at the low temperatures where superconductivity
can be observed much stronger than that expected from the
electron-hole asymmetry. Therefore, we concentrate on an
electron-hole symmetric theory in the remainder of this paper. We
employ the quasiclassical theory that provides a fair description of
inhomogeneous superconductivity both in equilibrium and
nonequilibrium systems. Moreover, for simplicity and also dictated
by many of the experiments, we concentrate on the diffusive limit.


\subsection{Usadel equations for proximity structures}
\label{sec:usadel}
\authors{Tero and Pauli}

Heterostructures composed of diffusive normal-metal or
superconducting wires in and out of equilibrium can be described
through the use of Usadel equations \cite{usadel70} for the Keldysh
Green's functions $\check{G}$. These equations are reviewed in many
references --- we cite here only a few of those
\cite{belzig99,virtanen04b} applying similar parametrization as
here. Written in the Nambu $\otimes$ Keldysh space, Usadel equation
is a nonlinear differential equation for a $4\times 4$ matrix,
\begin{equation}
D[\underline{\nabla}, \check{G} [\underline{\nabla}, \check{G}]]=\left[-i E +
\check{\Delta} + \check{\Sigma},\check{G}\right].
\label{eq:usadelmatrixform}
\end{equation}
Here $D$ is the diffusion constant, $E$ is the energy calculated from
the Fermi energy, $\check{\Delta}$ denotes the superconducting order
parameter and $\check{\Sigma}$ the self-energy for inelastic
scattering (mainly the part of electron--electron interaction not
described by $\check{\Delta}$ and electron--phonon scattering), for
spin-flip or spin-orbit scattering. In the presence of a magnetic
field, $\underline{\nabla}\equiv\nabla-ieA\hat{\tau}_3$ is the
gauge-invariant derivative including the vector potential $A$. In addition
to Eq.~\eqref{eq:usadelmatrixform}, $\check{G}$ satisfies the
normalization $\check{G}^2=\check{1}$, where $\check{1}$ is the
identity matrix.

In the diffusive limit, we implicitly assume that all the length
scales of the problem, including the superconducting coherence
length and the mean free paths for other types of scattering than
elastic, are much longer than the elastic mean free path. An example
of such other types of scattering is the spin-flip scattering,
described in the Born approximation by the
self-energy\cite{kopnin01}
\begin{equation}
\Sigma_{sf}=\frac{1}{2\tau_{sf}} \check{\tau}_3 \check{G}
\check{\tau}_3,
\end{equation}
where $\tau_{sf}$ is the spin-flip scattering time. This term is
included in the following analytic expressions, but omitted from the
numerics.

In the Keldysh space, Green's function has the form
\begin{equation*}
\check{G}=\begin{pmatrix} \hat{G}^R & \hat{G}^K\\0 &
\hat{G}^A\end{pmatrix},
\end{equation*}
where $\hat{G}^{R/A/K}$ denote the Retarded/Advanced/Keldysh
functions. The latter are $2\times 2$ matrices in the Nambu
particle-hole space.\footnote{Throughout the text, we employ the
notation where Keldysh matrices $\check{A}$ checked and Nambu
matrices $\hat{A}$ wear a hat. The Pauli matrices in Nambu space are
denoted by $\hat{\tau}_i$ and in Keldysh space by
$\check{\sigma}_i$, $i=1,2,3$.} Products of this type of matrices
yield similar matrices, without mixing the Keldysh parts into the
diagonal. Therefore, also the Usadel equation
\eqref{eq:usadelmatrixform} has a similar matrix structure.
Employing the normalization and the symmetry
$\hat{G}^A=-\hat{\tau}_3 \hat{G}^R \hat{\tau}_3$, we may parametrize
\begin{equation*}
\hat{G}^R=\cosh(\theta)\hat{\tau}_3+\sinh(\theta)(\cos(\chi) i
\hat{\tau}_2+\sin(\chi)i \hat{\tau}_1)
\end{equation*}
and
\begin{equation*}
\hat{G}^K=\hat{G}^R (f_L + f_T \hat{\tau}_3) - (f_L+f_T
\hat{\tau}_3)\hat{G}^A.
\end{equation*}
Here $\theta$ and $\chi$ are complex scalar parameters, roughly
describing the magnitude and phase of the pair amplitude,
respectively. In the Keldysh part, the additional parameters $f_L$
and $f_T$ are the longitudinal and transverse parts of the electron
distribution function. It can be shown \cite{kopnin01} that this
parametrization spans all the possible solutions of the
Keldysh-Usadel equations in non-magnetic systems.

Usadel equations for $\theta$ and $\chi$ are
\begin{subequations}
  \label{eq:spectral}
  \begin{gather}
  \begin{split}
    D \nabla^2 \theta =& -2i(E + i \Gamma_{\rm in}) \sinh\theta
    + \left(\frac{1}{\tau_{sf}}+\frac{v_S^2}{2D}\right)\sinh(2\theta) \\&+
    2 i |\Delta| \cos(\phi-\chi) \cosh(\theta)\,,
    \end{split}
    \label{eq:spectral1}
    \\
    \begin{split}
    \nabla\cdot(-v_S\sinh^2\theta) =& -2 i |\Delta| \sin(\phi-\chi) \sinh(\theta)\,,\\
    v_S \equiv& D(\nabla\chi - 2eA/\hbar)
    \,.
    \label{eq:spectral2}
    \end{split}
  \end{gather}
\end{subequations}
Here we assume that the superconducting order parameter is of the
form $\Delta=|\Delta|e^{i\phi}$. Note that in a proximity structure,
the superfluid velocity $v_S$ is position dependent. We include the
effect of weak inelastic scattering through a constant imaginary
part $\Gamma_{\rm in}$ of the energy \cite{schmid75}. In the
numerics, this is set to a small but finite positive value in order
to preserve the analytic structure of the Green's functions.

The kinetic equations for the distribution functions read
\begin{subequations}
  \label{eq:kin}
  \begin{align}
  \begin{split}
    D\nabla\cdot \hat\Gamma_T f &= (\nabla \cdot j_S) f_L + 2|\Delta| \Rc f_T, \\
    \hat\Gamma_T f &\equiv \DT\nabla f_T + \T\nabla f_L + \jS f_L
    \label{eq:kinjT}
    \,,
    \end{split}
    \\
    \begin{split}
    D\nabla\cdot \hat\Gamma_L f &= 0, \\
    \hat\Gamma_L f &\equiv \DL\nabla f_L - \T\nabla f_T + \jS f_T
    \label{eq:kinjL}
    \,,
    \end{split}
  \end{align}
\end{subequations}
where the kinetic coefficients are
\begin{subequations}
  \label{eq:coefficients}
  \begin{align}
    \DL &= \frac{1}{2}(1+|\cosh\theta|^2-|\sinh\theta|^2\cosh(2\Im[\chi])),
    \label{eq:DL}
    \\
    \DT &= \frac{1}{2}(1+|\cosh\theta|^2+|\sinh\theta|^2\cosh(2\Im[\chi])),
    \label{eq:DT}
    \\
    \T  &= \frac{1}{2}|\sinh\theta|^2\sinh(2\Im[\chi])),
    \label{eq:T}
    \\
    \jS &= \Im[-\sinh^2(\theta)\frac{v_S}{D}],
    \label{eq:js}
    \\
    \Rc &= \Im\left[-\cos(\phi-\chi) \sinh(\theta)\right].
    \label{eq:R}
  \end{align}
\end{subequations}
Inside a superconductor where the pair interaction parameter
$\lambda \neq 0$, the superconducting pair potential is obtained via
\begin{equation}\label{eq:selfcons}
\begin{split}
\Delta = \frac{\lambda}{4} \int dE [&
            (e^{i \chi}\sinh\theta + e^{i \chi^*} \sinh\theta^*) f_L\\&
                  - (e^{i \chi}\sinh\theta - e^{i \chi^*} \sinh\theta^*) f_T
        ]
        \end{split}
\end{equation}
Solving Eqs.~\eqref{eq:spectral}, \eqref{eq:kin} and
\eqref{eq:selfcons} we get the observables, for example the charge
and energy current densities given by
\begin{align}
  \label{eq:currents}
  J_c = -\frac{\sigma}{2 e} \int_{-\infty}^{\infty}\dd{E} \hat\Gamma_T f \,, \quad
  J_E = \frac{\sigma}{2 e^2} \int_{-\infty}^{\infty}\dd{E} E\,\hat\Gamma_L f \,.
\end{align}
In most of the text below, we assume that the superconductors are
bulky reservoirs, such that the self-consistency equation
\eqref{eq:selfcons} can be ignored. We rather concentrate on the
phenomena taking place in normal-metal wires close to the
superconductors. In those wires, we assume $\lambda=0$, and thereby
also $\Delta=0$. This simplifies the resulting equations.

\subsection{Interfaces and terminals}
\label{sec:interfaces}

Usadel equation holds within the wires where changes in the
parameters take place slowly compared to the mean free path. At
interfaces, it has to be supplemented by boundary conditions.
Initially, these were derived for a general quasiclassical Green's
function by Zaitsev \cite{zaitsev84}. For the diffusive case, the
general boundary conditions were solved by Nazarov \cite{nazarov99}.
They read
\begin{equation}
\begin{split}
\check{I}_L=\check{I}_R &= \frac{2e^2}{\pi\hbar}
  \sum_n \frac{\tau_n [\check{G}_L,\check{G}_R]}{
    4 - \tau_n(\{\check{G}_L,\check{G}_R\} - 2)},\\
    \check{I}_i&\equiv \sigma_i A_i \check{G}_i \nabla \check{G}_i \cdot
    \uvec{n},
 \end{split}
\end{equation}
where $\sigma$ and $A$ are the normal-state conductivity and the
cross section of the wires next to the interface, subscript $L/R$
denote left/right from the interface, and $\uvec{n}$ is the unit
vector perpendicular to the interface, pointing to the right. The
interface is characterized by the set $\{\tau_n\}$ of transmission
eigenvalues. Note that the resulting expression for $\check{I}$ is
linear in the electron distribution functions $f$, due to the
Keldysh block structure of the Green's functions. In what follows,
we assume that the normal-state conductance $G_I=2e^2 \sum_n
\tau_n/h$ for each interface is large, such that their effect can be
neglected. However, the arguments on the general symmetries of the
thermoelectric coefficients are independent of this assumption.


In addition to having the correct boundary conditions for interfaces,
one needs also to describe the behavior of the Green's function
$\check{G}$ inside different types of terminals. A typical assumption
is that the Green's functions obtain their bulk values very close to
the interface between a wire and a terminal.  Essentially this means
that the specific resistance (both charge and thermal) of the
terminals should be much smaller than that of the mesoscopic region
under study. Experimentally this is realized by making the cross
section of especially the normal-metal terminals much larger than that
of the wires.

Inside superconductors for energies $E<\abs{\Delta}$ all quantities
except $f_L$ relax to their bulk values within distances comparable
to the coherence length $\xi_0 = \sqrt{\hbar D/(2 \Delta)}$.
Moreover, for the L-mode, the Andreev reflection boundary condition
$\hat{\Gamma}_Lf=0$ applies at these energies. These details of NS
interfaces one can usually describe by increasing the effective
length \cite{likharev79} of the normal-metal wires in contact to
superconductors by an amount comparable to $\xi_0$. However, at
energies $E>\abs{\Delta}$, nonequilibrium in $f_T$ and $f_L$ may
persist to greater distances.  This charge and energy imbalance is
limited by inelastic relaxation processes, and for the charge mode
in the diffusive limit, by the decoherence induced by a flowing
supercurrent or spin-flip scattering.  (See for example
Refs.~\cite{schmid75,tinkham}.)

For temperatures or voltages of the order or larger than $\Delta$,
we hence have to pay some attention to a proper treatment of
superconductors, especially superconducting loops (for an example,
see Fig.~\ref{fig:setup}) with length $L_L$, cross section $A_L$ and
normal-state conductivity $\sigma_L$. Assume such a loop is
connected to a normal-metal wire with length $L_w$, cross section
$A_w$ and normal-state conductivity $\sigma_w$. When compared to the
superconductor, the latter is described by an effective length
$L_w'=L_w \sigma_L A_L/(\sigma_w A_w)$ to account for the
differences in the specific resistance. Furthermore, assume an
energy relaxation length $L_E$ inside the superconductor. We then
have three practically important limits: a) $L_E \ll L_L, L_w'$, b)
$L_w' \ll L_L \ll L_E$ and c) $L_w' \ll L_E \ll L_L$. In the first
case, the relaxation in the superconductors is fast, and we may
assume that $f_L(E > \abs{\Delta})$ and $f_T(E>\abs{\Delta})$
acquire their bulk values immediately at the superconducting
interface.  In the case b), the normal-state resistance of the loop
is much higher than that of the normal-metal wires, so that the
proper boundary condition is the vanishing of quasiparticle current
to the superconductors. For the case c), we again get a vanishing of
the quasiparticle charge current, but the energy current will depend
on the details of inelastic relaxation in the superconductor.

We see no way to formulate exact mathematical boundary conditions for
the limits b) and c) above---they in principle require the solution of
the Usadel equation inside the superconductor. One attempt to
approximate the case b) in a way consistent with the Onsager symmetry
is described in Sec.~\ref{subs:loops}. It captures most of the
essential physics of this problem, i.e., taking into account the
finite charge and thermal resistance of the loop at high temperatures.

\subsection{Effects left out}
\label{subs:effectsleftout}
\authors{Tero}

There are two more practically important self-energies that were not
included in the above description: those related to
electron--electron and electron--phonon interactions,
$\check{\Sigma}_{\rm e-e}$ and $\check{\Sigma}_{\rm e-ph}$. These
two have a few distinct characteristics compared to the included
scattering mechanisms (mainly elastic and spin-flip scattering):
\begin{itemize}
\item They are {\it inelastic} scattering mechanisms, i.e., they
lead to the non-conservation of spectral currents. This is why these
should be taken into account similarly as the self-consistency
relation, Eq.~\eqref{eq:selfcons}. However, electron-electron
scattering conserves the total energy and charge current, whereas
electron-phonon scattering conserves only the charge current.
\item These scattering mechanisms provide both dephasing and energy
relaxation, i.e., both their Retarded/Advanced and Keldysh parts are
finite.
\item Similarly to the self-consistency relation, these scattering
mechanisms make the equations for the Retarded/Advanced functions
depend on the distribution functions $f_L$ and $f_T$.
\end{itemize}
The self-energies for these scattering mechanisms in the presence of
superconductivity are detailed in Ref.~\cite{kopnin01}.

Furthermore, as we concentrate only on the diffusive limit, we
neglect effects related to different types of elastic scattering.

\section{Supercurrent spectrum and nonequilibrium electron energy distribution
function} \authors{Tero} \label{sec:scspectrum}

The presence of the supercurrent-induced terms $\jS$ and $\T$ in
Eq.~\eqref{eq:kin} leads to the finite thermoelectric effects
described in Sec.~\ref{sec:thermocoefs}. But before engaging to
their discussion, let us take a look at the spectral supercurrent
$\jS$ and how its form can be employed together with a
nonequilibrium distribution function to tune the supercurrent
flowing in a Josephson junction, or alternatively, to modify the
energy distribution function.

\subsection{Spectral supercurrent}

If a phase-coherent normal-metal wire is sandwiched between two
superconductors, Andreev reflection at each NS interface results
into a formation of Andreev bound states \cite{andreev66,kulik70}.
In the case of a clean normal metal, these bound-state energies
depend on the phase difference $\varphi$ between the superconducting
contacts, the traversal time $d/v_F$ through the normal-metal region
of length $d$, and the transparency $\tau$ of the NS interface. For
a junction much longer than the superconducting coherence length,
the bound-state energies are \cite{kopnin06}
\begin{equation}
\varepsilon_n^{\pm} = \pm \frac{\hbar v_F}{d}\left(\arcsin
\sqrt{\tau^2
\cos^2\left(\frac{\varphi}{2}\right)+(1-\tau^2)\sin^2(\alpha)}+n
\pi\right).
\end{equation}
Here $\alpha=k_F d+\delta$ is the dynamical phase gathered within
traversal through the junction, $\delta$ depending on the phase shift
at the interface. The characteristic property of these bound states is
that they carry an amount of supercurrent proportional to the phase
derivative of the bound-state energy. Therefore, we can define a
"spectral supercurrent" via
\begin{equation*}
\jS \sim \sum_m \frac{\partial \varepsilon_m^{\pm}}{\partial \varphi}
\delta(E-\varepsilon_m).
\end{equation*}
In the clean limit $\jS$ would hence contain a sequence of delta
peaks. In the diffusive limit on which we concentrate in this paper,
the Andreev state spectrum becomes continuous as disorder makes rise
to a distribution of transparencies and times of flights. In this
case, $\jS$ can be calculated by solving Eqs.~\eqref{eq:spectral}
with proper boundary conditions. Its behavior in different limits is
detailed in Refs.~\cite{wilhelm98,heikkila02scdos}. An example of
$\jS(E)$ specific to the geometries considered in this paper is
presented in Fig.~\ref{fig:spectralsupercurrent}.

If no dc voltage between the superconductors is applied, the
supercurrent between them is obtained from Eq.~\eqref{eq:currents},
\begin{equation}
I_S = \frac{\sigma A}{2 e} \int_{-\infty}^\infty \dd{E} f_L(E)
\jS(E).
\end{equation}
Attaching normal-metal terminals to the wire allows one to tune the
energy distribution function $f_L(E)$, and thereby the supercurrent
\cite{vanwees91,wilhelm98,yip98,volkov95}. Such nonequilibrium
supercurrent was experimentally demonstrated around the turn of the
century by many groups
\cite{baselmans99,baselmans01,huang02,schapers98,kutchinsky99,shaikhaidarov00}.
One of the most interesting features of these experiments is the
possibility to take the junction into the $\pi$-state, where the
ground state of the junction corresponds to a phase difference of $\pi$
between the contacts, and the supercurrent for a given phase
difference is reversed compared to the usual
0-state.\cite{baselmans99,baselmans02}. This $\pi$-state occurs when
the distribution function $f_L$ weighs the negative part of the
supercurrent spectrum more than the positive part (c.f.,
Fig.~\ref{fig:spectralsupercurrent}).

\begin{figure}\centering
  \includegraphics{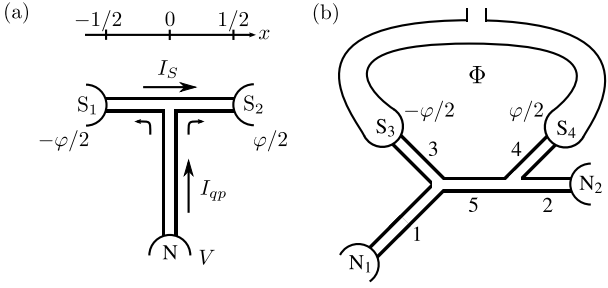}
  \caption{\label{fig:setup}
    (a)
    Three-probe structure consisting of two superconducting terminals
    and one normal-metal terminal. The phase difference $\varphi$ between
    the superconducting terminals drives supercurrent $I_S$, and the voltage
    bias $V$ in the normal terminal drives quasiparticle current $I_{qp}$.
    (b) Andreev interferometer, consisting of a
    superconducting loop and two normal-metal terminals, connected by
    5 normal-metal wires.  The magnetic flux $\Phi$ threading the loop
    controls the superconducting phase difference
    $\varphi\equiv2\pi\Phi/\Phi_0$.  We take the relative lengths of
    the wires to be $L_j/L_{SNS} = \frac{2}{3}, \frac{1}{3},
    \frac{1}{3}, \frac{1}{3}, \frac{1}{3}$ and assume the wires to
    have the same cross-sectional area $A$ and conductivity $\sigma$.  In the
    numerics, we assume the wires to be quasi-one-dimensional,
    $\sqrt{A}\ll{}L$. The absolute size of the system controls
    the characteristic Thouless energy scale $E_T=\hbar D/L_{SNS}^2$,
    with $L_{SNS}=L_3+L_4+L_5$.
  }
\end{figure}

\begin{figure}\centering
  \includegraphics{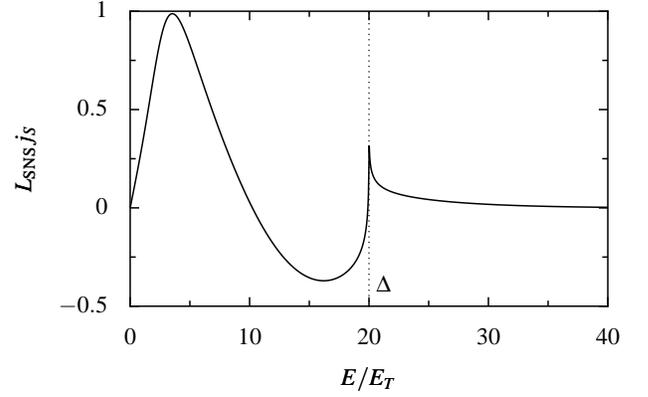}
  \caption{\label{fig:spectralsupercurrent} Spectrum of the
    supercurrent in wires 3, 4, 5 in the structure of
    Fig.~\ref{fig:setup}(b), for phase difference $\varphi=1.6$, and
    superconducting gap $\abs{\Delta} = 20E_T$.  }
\end{figure}

\subsection{Driving a nonequilibrium energy distribution with
supercurrent}
\label{sec:drivingf}

\begin{figure}\centering
  \includegraphics{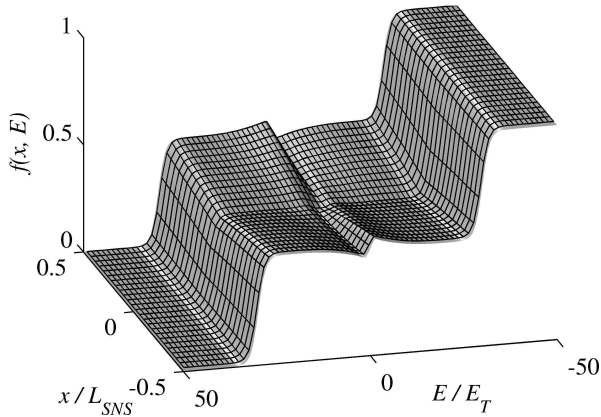}
  \caption{\label{fig:drivenf} Electron distribution function $f(x,E)
    = \frac{1}{2}[1-f_T(x,E)-f_L(x,E)]$ between the two
    superconducting terminals in Fig.~\ref{fig:setup}a.  The bias
    voltage is chosen $V=30E_T/e$, temperature is $T=1E_T/k_B$, and
    $\abs{\Delta}\gg{}E_T$ and $\varphi=\pi/2$ are assumed.  The
    low-energy ($E\sim{}E_T$) perturbation in $f$ arises from the L/T
    mixing in the proximity effect, and the $2V$-step from the Andreev
    reflection, see text.  }
\end{figure}

Let us consider the solution to the kinetic equations \eqref{eq:kin}
in a three-probe system depicted in Fig.~\ref{fig:setup}(a). The two
superconducting terminals are assumed to be at zero potential,
whereas the normal-metal terminal is at potential $V$. For
simplicity, let us assume the system left-right symmetric. In this
case, the following symmetries apply inside the horizontal wire:
\begin{align*}
\jS(\varphi)&=-\jS(-\varphi)\\
\T(\varphi,x)&=-\T(-\varphi,x)=-\T(\varphi,-x)\\
\DT(\varphi,x)&=\DT(-\varphi,x)=\DT(\varphi,-x)\\
\DL(\varphi,x)&=\DL(-\varphi,x)=\DL(\varphi,-x)
\end{align*}
In the vertical wire, we hence have $\jS=\T=0$, and the kinetic
equations for $f_T$ and $f_L$ are decoupled. Let us now try to solve
for $f_L(x)=f_L^0+\delta f_L(x)$ in the horizontal wire. Here
$f_L^0=[\tanh((E+eV)/(2 k_B T))-\tanh((E-eV)/(2k_B T))]/2$ is the
longitudinal distribution in the normal terminal. Using the fact
that for $\abs{E} < \abs{\Delta}$, $\hat{\Gamma}_L f=0$ throughout
the normal-metal system, we can find an exact solution for these
energies:
%
\begin{equation}
\delta f_L(x)=\int_0^x dx' \frac{\T(x')}{\DL(x')} (\partial_x
f_T)_{x=x'}-\jS \int_0^x dx' \frac{f_T(x')}{\DL(x')}.
\label{eq:flsol}
\end{equation}
This solution can now be substituted to Eq.~\eqref{eq:kinjT}. The
latter yields a second-order linear differential equation for $f_T$,
independent of $f_L$.
From the full numerical solution we can find that the proximity
corrections to $f_T$ are relatively small compared to those in
$\delta f_L$. Therefore, let us neglect those corrections and solve
Eq.~\eqref{eq:kinjT} in the incoherent limit $\DT=1$, $\T=\jS=0$. In
this case we get $f_T(x)=(1-\frac{2|x|}{L_{SNS}})f_T^c$, where
$f_T^c=\rho_A f_T^0$
is the transverse function at the crossing point $x=0$. Here
$f_T^0=[\tanh((E+eV)/(2 k_B T))+\tanh((E-eV))/(2 k_B T)]/2$ is the
boundary condition for $f_T$ in the normal reservoir,
$\rho_A=(\sigma_V A_V L_{SNS})/(\sigma_V A_VL_{SNS}+4\sigma_{SNS}
A_{SNS} L_V)$, and $A_{SNS/V}$ are the cross sections and
$\sigma_{SNS/V}$ the normal-state conductivities of the horizontal
and vertical wires, respectively. Substituting this solution to
Eq.~\eqref{eq:flsol} finally yields
\begin{equation}
\delta f_L(x)=-f_T^0 \rho_A\left[\frac{2}{L_{SNS}}\int_0^x dx'
\frac{\T(x')}{\DL(x')} +\jS \int_0^x dx'
\frac{1-\frac{2x}{L_{SNS}}}{\DL(x')}\right]. \label{eq:flsol2}
\end{equation}
We thus find that the supercurrent controls the antisymmetric part
of the distribution function: for a vanishing phase gradient across
the wire, $\delta f_L=0$. For $k_B T \ll e V$, $f_T^0$ defines a
window of energy $E \in [-eV,eV]$ in which the correction is finite
(there, $f_T^0 \approx 1$, whereas $f_T^0 \approx 0$ for $\abs{E}
> \abs{eV}$). Close to the crossing point $x=0$, $\DL \approx 1$, and the
energy dependence of $\delta f_L(x)$ reflect directly those of
$\T(x)$ and $\jS$. Close to the NS interface $x \rightarrow \pm
\frac{1}{2}$, $\DL$ tends to zero, and both of the terms in
Eq.~\eqref{eq:flsol2} diverge. However, their sum stays finite and
the remaining part is roughly proportional to the spectral
supercurrent $\jS$. The full distribution function $f(E,x)$ in the
horizontal wire is plotted in Fig.~\ref{fig:drivenf} for one example
value of the phase difference.  The supercurrent-induced changes in
the nonequilibrium distribution function were recently measured
\cite{crosser06}, and the results were in a fair agreement with the
theory sketched above.

The longitudinal distribution function (the energy mode) $f_L$
describes the response of the electron system to changes in the
temperature \cite{tinkham96}. In this way, the above changes in
$f_L$ can be understood as supercurrent-driven modifications in the
local temperature \cite{heikkila03}: due to the antisymmetry of
$\delta f_L(x)$ about the crossing point $x=0$, one of the
horizontal arms heats up, and another one cools down. Such a setup
thus resembles a Peltier-like system. However, in this case one has
to deal with an effective temperature $T_{\rm eff}$ (for its
definition, see Refs.~\cite{heikkila03,heikkilathesis}), and it
turns out that for this symmetric system the increase in $T_{\rm
eff}$ due to the Joule heating is always larger than the changes due
to the supercurrent. Both of these issues are settled below when
considering the properties of an arbitrarily shaped four-terminal
interferometer.

\section{Multi-terminal thermoelectric coefficients}
\label{sec:thermocoefs}

\authors{Pauli}

In this section, we apply the theory formulated in
Sec.~\ref{sec:thermoelectrictransport} to calculate the
multiterminal transport coefficients defined in
Eq.~\eqref{eq:linearresponseNS}. Main emphasis is on the appearance
of thermoelectric effects, which originate from the same mixing of
the L and T modes that in Fig.~\ref{fig:drivenf} modifies the shape
of the electron distribution function.
Below, we calculate all thermoelectric transport coefficients in the
same example setup shown in Fig.~\ref{fig:setup}(b), a typical
instance of an Andreev interferometer. The interference effects due
to superconductivity are tuned by the magnetic flux $\Phi$ threading
the superconducting loop, which adjusts the superconducting phase
difference $\varphi$, and observed by measuring various transport
properties of the wire between the two normal terminals.  We assume
here the structure to be left--right asymmetric, not to miss certain
effects that vanish in completely symmetric structures.

\subsection{Spectral thermoelectric matrix}
\label{sec:Lmatrix}
\authors{Tero and Pauli}


Based on the above discussion, one could examine transport in
proximity structures simply by solving the Usadel equations
numerically and evaluating the current--bias relation for all
necessary values of temperatures and voltages at the reservoirs.
However, for the proximity effect, it is possible to separate the
biases from the full non-linear response of the circuit by making only
mild assumptions.


First, one can note that the only part of the above equations that
is nonlinear in the electron distribution functions $f$ is the
self-consistency equation~\eqref{eq:selfcons}. Neglecting it is
often a good approximation if the terminals are large compared to
the rest of the system. Disregarding Eq.~\eqref{eq:selfcons}, the
linearity in $f$ directly allows one to write the charge and thermal
current $I_c^i$ and $I_E^i$ entering a given reservoir $i$ as a
linear combination of the distribution functions $f_{\alpha}^j(E)$
in all reservoirs: \cite{virtanen07}
\begin{subequations}\label{eq:Lmatrix}
\begin{align}
  I_{c}^{i} &= \int_{-\infty}^\infty \dd{E}\,
             \sum_{\beta j} \tilde{L}_{T\beta}^{ij}(E) f_{\beta}^{j}(E)
  \,,
  \\
  I_{E}^{i} &= \int_{-\infty}^\infty \dd{E}\,E\,
             \sum_{\beta j} \tilde{L}_{L\beta}^{ij}(E) f_{\beta}^{j}(E)
  \,.
\end{align}
\end{subequations}
Similar decomposition has been used in the literature mostly for
describing charge transport. \cite{volkov96,courtois99} Below, we
call the set of functions $\tilde{L}_{\alpha\beta}^{ij}(E)$ the
spectral thermoelectric matrix, because the thermoelectric
linear-response coefficients are related to it in a natural way:
\begin{subequations}\label{eq:linearresponseL}
\begin{align}
  L_{11}^{ij}&=\frac{1}{2k_BT}\int\dd{E}
  \tilde{L}_{TT}^{ij}(E)\sech^2\left(\frac{E}{2k_BT}\right) \,,
  \\
  L_{21}^{ij}&=\frac{-1}{2k_BT}\int\dd{E}
  E\,\tilde{L}_{LT}^{ij}(E)\sech^2\left(\frac{E}{2k_BT}\right) \,,
  \\
  L_{12}^{ij}&=\frac{-1}{2k_BT}\int\dd{E}
  E\,\tilde{L}_{TL}^{ij}(E)\sech^2\left(\frac{E}{2k_BT}\right) \,,
  \\
  L_{22}^{ij}&=\frac{1}{2k_BT}\int\dd{E}
  E^2\,\tilde{L}_{LL}^{ij}(E)\sech^2\left(\frac{E}{2k_BT}\right)
  \,.
\end{align}
\end{subequations}
In principle, the functions $\tilde{L}^{ij}_{\alpha\beta}(E)$ are a
generalization of the plain linear-response coefficients.


The matrix element $\tilde{L}^{ij}_{\alpha\beta}(E)$ can be defined
explicitly as the $\alpha$-mode current flowing in terminal $i$
in response to a $\beta$-mode unit excitation in terminal $j$,
at energy $E$:
\begin{align}\label{eq:Ldefinition}
  \tilde{L}^{ij}_{\alpha\beta}(E) \equiv
  \int_{\mathcal{S}_i} \dd{\mathcal{S}} \,
  \uvec{n} \cdot \hat\Gamma_{\alpha} \psi^{j,\beta}(E)
  \,,
\end{align}
where $\mathcal{S}_i$ is the surface of the $i$:th terminal and
$\uvec{n}$ the corresponding normal vector. The two-component
characteristic potential
$\psi^{j,\beta}=(\psi^{j,\beta}_T,\psi^{j,\beta}_L)$ is assumed to
satisfy the kinetic equations together with their boundary conditions,
with the distribution function $f_\alpha^i$ in each terminal
replaced by $\delta_{\alpha\beta}\delta_{ij}$.

\begin{figure}\centering
  \includegraphics{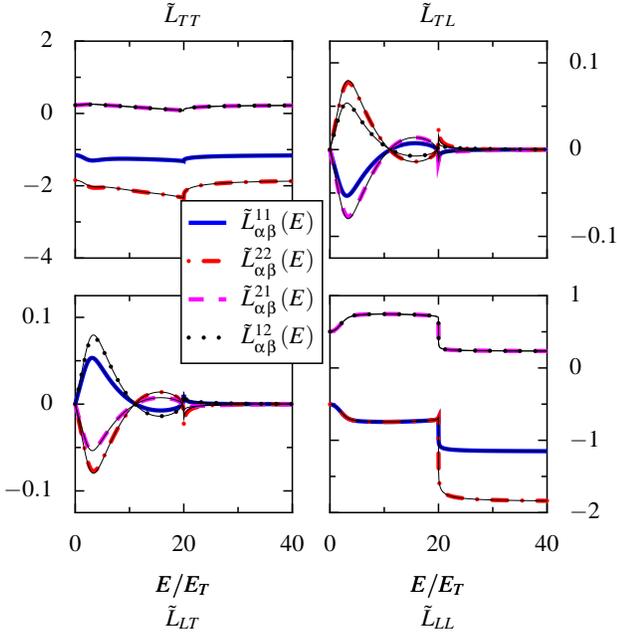}
  \caption{\label{fig:Lmatrix} Elements of the spectral thermoelectric
    matrix $\tilde{L}_{\alpha\beta}^{ij}(E)$ associated with the
    normal terminals, $i,j=1,2$, in the structure of
    Fig.~\ref{fig:setup}(b).  Phase difference is assumed to be
    $\varphi=1.6$ and the superconducting gap $\abs{\Delta}=20E_T$.  }
\end{figure}

\begin{figure}\centering
  \includegraphics{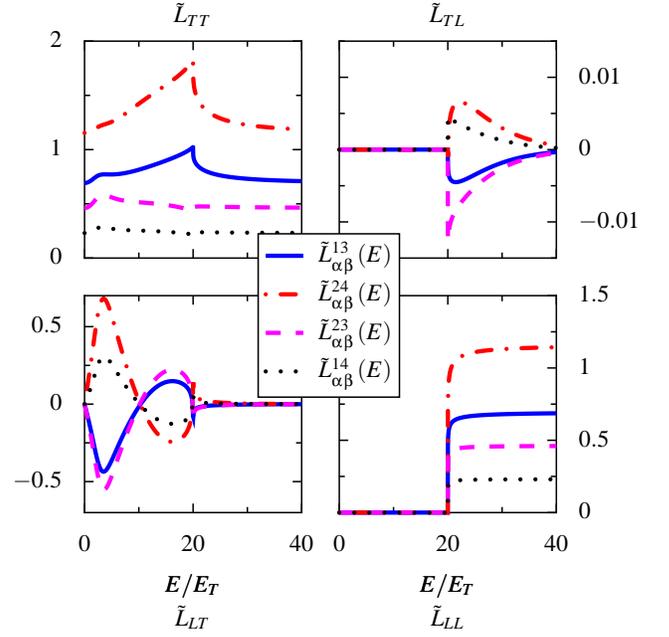}
  \caption{\label{fig:LSmatrix}
    Elements of the spectral
    thermoelectric matrix $\tilde{L}_{\alpha\beta}^{ij}(E)$
    associated with excitations in the superconductor, $i=1,2$, $j=3,4$.
    Assumptions are as in Fig.~\ref{fig:Lmatrix}.
  }
\end{figure}

Examples of the energy dependence of the
$\tilde{L}_{\alpha\beta}^{ij}(E)$ functions for the four-terminal setup
in Fig.~\ref{fig:setup}(b) are shown in Figs.~\ref{fig:Lmatrix},
\ref{fig:LSmatrix}.  The two characteristic energy scales for these
coefficients are, similarly as for the spectral supercurrent, the
Thouless energy $E_T = \hbar D/L_{SNS}^2$ and the superconducting
energy gap $\Delta$.  Note that since our theory is limited to static
situations, only $L$-mode (temperature) bias can be applied to the
superconductors if they are at internal equilibrium --- for many
phenomena, the coefficients in Fig.~\ref{fig:Lmatrix} are more
relevant than those in Fig.~\ref{fig:LSmatrix}. However, a
nonequilibrium $T$-mode bias could be generated within the static
model by inducing charge imbalance in the superconductors, for example
by injecting current from additional normal-metal junctions.


Semi-analytical expressions for the coefficients
$\tilde{L}_{\alpha\beta}^{ij}(E)$ can be found by solving Eqs.~\eqref{eq:kin}
up to first order in $j_S$ and $\T$. In systems that can be considered
as a circuit of quasi-1D wires, this leads to a circuit theory for the
distribution functions. Between two nodes with distribution functions
$f^1=(f_T^1, f_L^1)$ and $f^2=(f_T^2, f_L^2)$, one finds an expression
for the spectral currents
\begin{subequations}\label{eq:linearization}
  \begin{align}
    \hat{\Gamma} f
    &\simeq
    (\hat{M}^{-1} - t i\hat{\tau}_2 + \frac{\gamma j_S}{2} i\hat{\tau}_2) (f^2 - f^1)
    \\\notag&\qquad+ \frac{j_S}{2}\hat{\tau}_1(f^2 + f^1)
    + \mathcal{O}(j_S^2 + \T^2)
    \,,
    \intertext{where $\hat{\tau}_1$ and $\hat{\tau}_2$ are Nambu spin matrices, and}
    \hat{M}&\equiv\mathrm{diag}(M_T, M_L) \,,
    \quad M_\alpha \equiv \int_0^L\dd{x}\mathcal{D}_\alpha(x)^{-1}
    \,,
    \\
    t &\equiv \int_0^L\dd{x}\T(x)\frac{\DL(x)^{-1}\DT(x)^{-1}}{M_LM_T}
    \,,
    \\
    \gamma &\equiv
    \int_0^L\dd{x}\!\int_0^L\dd{x'} \sgn(x-x')
    \frac{\DL(x)^{-1}\DT(x')^{-1}}{M_LM_T}
    \,.
  \end{align}
\end{subequations}
If node 1 (or node 2) is at a clean interface to a bulk superconductor
at $E<\abs{\Delta}$, one can use the asymptotic behavior $\DL(x)={\rm
  const.}\times{}x^2 + {\cal O}(x^3)$, $\T(x)=j_S x+{\cal O}(x^2)$ to
find $M_L^{-1}=0$, $t=0$, $\gamma=\pm1$.  Using conservation of the
spectral current $\hat{\Gamma}f$ at the nodes and suitable boundary
conditions, one can in this way find an approximation to
$\tilde{L}_{\alpha\beta}^{ij}(E)$ for any given circuit.  The
quality of this approximation is usually quite good---in
Fig.~\ref{fig:Lmatrix} such approximations are shown with black
lines, which almost coincide with the numerical results.  However,
the spectral equations need still to be solved to find out the
proximity-modified diffusion constants $D\mathcal{D}_\alpha$, $\T$
and the spectral supercurrent $j_S$.


\subsection{Symmetry relations}
\label{sec:symmetry}
\authors{Pauli}


As discussed in Section~\ref{sec:normaltransport}, the normal-state
thermoelectric transport coefficients are usually coupled together
by Onsager's reciprocal relation
$L^{ij}_{\alpha\beta}(B)=L^{ji}_{\beta\alpha}(-B)$ under the
reversal of the magnetic field. The question now is: do the
thermoelectric coefficients induced by the proximity effect follow
this same relation, and what else can we say about their symmetries.
In the framework of scattering theory, it turns out that the Onsager
reciprocity applies also in hybrid normal--superconducting systems.
\cite{hartog96,claughton96} Moreover, within the Usadel theory, it
has been shown that the off-diagonal coefficients $L_{12}$, $L_{21}$
are always odd functions of the magnetic field $B$, whereas the
diagonal coefficients $L_{11}$, $L_{12}$ are even.
\cite{virtanen07,virtanen04b,seviour00} Below, we review the
symmetries present in the Usadel framework.


That a form of Onsager's reciprocal relation applies for the Usadel
model can be seen from the structure of the kinetic equations
\eqref{eq:kin} and symmetries of the coefficients
\eqref{eq:coefficients} under the reversal of the magnetic fields
$B$ (i.e., change of sign in the vector potential $A$ and the
superconducting phases $\phi$, $\chi$).  The crucial observation is
that the differential operator $\hat{\mathcal{O}}$ in the kinetic
equations \eqref{eq:kin}, $\hat{\mathcal{O}}f = 0$, is related to
its operator adjoint by \cite{virtanen07}
\begin{align}\label{eq:adjointrelation}
  \hat{\mathcal{O}}(B)^\dagger
  &= (-\nabla)\cdot\mat{\DT & -\T \\ \T & \DL}(-\nabla)
    + (-\nabla)\cdot j_S\hat{\tau}_1
    \\\notag
    &\quad
    - \mat{ 2\abs{\Delta}{\cal R} & -\nabla \cdot j_S \\ 0 & 0  }
    \\\notag
  &= \hat{\mathcal{O}}(-B) \,.
\end{align}
Here we exploited the symmetries
$\mathcal{D}_\alpha(-B)=\mathcal{D}_\alpha(B)$, $\T(-B)=-\T(B)$,
${\cal R}(-B)={\cal R}(B)$, and $j_S(-B)=-j_S(B)$ of the kinetic
coefficients \eqref{eq:coefficients}. From the above relation, it
follows that for any two-component functions $\phi$, $\rho$,
\begin{align}\label{eq:onsagerpartialinteg}
  \int_{\Omega}\dd{\mathcal{V}}[\rho^\dagger \hat{\mathcal{O}} \phi - \phi^\dagger \hat{\mathcal{O}}^\dagger \rho]
  = \int_{\partial\Omega}\dd{\mathcal{S}} \uvec{n} \cdot J \,,
\end{align}
where the flux
$J=\rho^\dagger\hat\Gamma(B)\phi-\phi^\dagger\hat\Gamma(-B)\rho-j_S\rho^\dagger\hat{\tau}_1\phi$
is what is left over from the integration by parts on the left-hand
side. Especially, this flux is conserved when $\phi$ satisfies the
kinetic equations for $+B$, and $\rho$ for $-B$. Making now use of
the functions applied in Eq.~\eqref{eq:Ldefinition} and substituting
$\phi=\psi^{j,\beta}(+B)$, $\rho=\psi^{i,\alpha}(-B)$, the
conservation of $J$ in the volume $\Omega$ of the structure implies
\begin{gather}\label{eq:implicitonsager}
  \begin{split}
  0
  &= \int_{\Omega}\dd{\mathcal{V}}[\rho^\dagger \hat{\mathcal{O}} \phi - \phi^\dagger \hat{\mathcal{O}}^\dagger \rho]
  = \int_{\partial\Omega}\dd{\mathcal{S}}\uvec{n}\cdot J \\
  &= \int_{\mathcal{S}_i}\dd{\mathcal{S}} \uvec{n}\cdot\hat\Gamma_{\alpha}(B) \phi
  - \int_{\mathcal{S}_j}\dd{\mathcal{S}} \uvec{n}\cdot\hat\Gamma_{\beta}(-B) \rho
  \,,
  \end{split}
\end{gather}
when both $i$ and $j$ refer to normal terminals. In this case the
last term in $J$, being proportional to $j_S$, vanishes on the
terminal surfaces $\mathcal{S}_i$ and $\mathcal{S}_j$. Other terms
vanish due to the boundary conditions assumed for the $\psi$
functions. Comparing this result to Eq.~\eqref{eq:Ldefinition}, one
finds for $i$, $j$ referring to the normal terminals
\begin{align}\label{eq:onsagersymmetry}
  \tilde{L}_{\alpha\beta}^{ij}(E,B)=\tilde{L}_{\beta\alpha}^{ji}(E,-B)
  \,,
\end{align}
which is a form of Onsager's reciprocal relation.

A second class of symmetries arises from the way the coefficients
$\tilde{L}_{\alpha\beta}^{ij}(E)$ were defined in Eq.~\eqref{eq:Lmatrix}.
Namely, we must require that
\begin{subequations}\label{eq:Lsymmetriesforced}
\begin{gather}
  \sum_{j} \tilde{L}^{ij}_{LL}(E) = 0 \,,\\
  \sum_{j} \tilde{L}^{ij}_{TL}(E) = 0 \qquad\text{for normal terminal $i$,}
\end{gather}
\end{subequations}
so that no net energy current flows to any terminal at equilibrium for
any temperature, and that the same applies for the charge current
entering the normal terminals.

The third symmetry relation is important for the thermoelectric
effects, and is specific to the quasiclassical theory. Namely, if
Green's function $\check{G}_1$ is a solution to the Usadel equation
for vector potential $A$ and self-energy
$\check{X}_1[\check{G}_1]$,
\begin{equation}
  [\nabla - ieA\hat{\tau}_3, \check{G}_1[\nabla - ieA\hat{\tau}_3, \check{G}_1]]=[\check{X}_1[\check{G}_1], \check{G}_1]
  \,,
\end{equation}
then, the electron--hole transformed Green's function
$\check{G}_2\equiv-\hat{\tau}_1\check{G}_1\hat{\tau}_1$ is a solution to the same
equation for $-A$ and self-energy
\begin{equation}\label{eq:transformedselfenergy}
\check{X}_2[\check{G}_2]=-\hat{\tau}_1\check{X}_1[-\hat{\tau}_1\check{G}_2\hat{\tau}_1]\hat{\tau}_1.
\end{equation}
For $\check{X}_1[\check{G}] = -iE\hat{\tau}_3 +
\hat{\Delta}[\check{G}] +
\frac{1}{2\tau_{sf}}\hat{\tau}_3\check{G}\hat{\tau}_3$ used above, we
note that $\check{X}_2(B) = \check{X}_1(-B)$ --- the two functionals
coincide. Hence, the transformed Green's function describes the same
physical situation, but with an inverted magnetic field. Since
electric potentials and charge currents also change sign under this
transformation, one finds that \cite{virtanen04b,virtanen07}
\begin{gather}\label{eq:Lsymmetryqcl}
  \tilde{L}^{ij}_{\alpha\beta}(E,-B)
  = (-1)^{1-\delta_{\alpha\beta}}\tilde{L}^{ij}_{\alpha\beta}(E,B)
  .
\end{gather}
This symmetry makes the off-diagonal thermoelectric coefficients odd
functions of the applied magnetic field, which is not in agreement
with all experiments.  We discuss this discrepancy in more detail in
Section~\ref{sec:discussion} and in the Appendix.

\subsubsection{Charge imbalance in superconducting loops}
\label{subs:loops}

Below, one of the aims is to model qualitative features of charge
imbalance in superconducting loops (see Sec.~\ref{sec:interfaces} and
Fig.~\ref{fig:setup}) without solving the Usadel equations inside
superconductors. For this, we need some effective boundary conditions
to enforce at the NS interfaces instead of the usual terminal
assumption [case a) in Sec.~\ref{sec:interfaces}]. Consider a
superconducting loop with a large normal-state resistance but long
inelastic relaxation length [case b) in Sec.~\ref{sec:interfaces}].
Deep in the superconductor, we then assume that the charge current is
carried only as supercurrent with the (BCS) spectral density
$j_S\propto\delta(E-\abs{\Delta})$.  Due to the large resistance, we
can also assume $\hat{\Gamma}_L\phi=0$ and $\hat{\Gamma}_T\phi=0$ for
$E\ne\abs{\Delta}$, for any solution $\phi$ of the kinetic equations.
Near the interface, supercurrent conversion occurs and the
$\delta$-peak in $\hat{\Gamma}_T\phi$ broadens, which needs to be
handled correctly to preserve Onsager reciprocity.  Equation
\eqref{eq:onsagerpartialinteg} defines a flux $J$ that is conserved in
the superconductor. By our assumptions, $J=0$ deep in the
superconductor, for $E\ne\abs{\Delta}$.  The exact solution $f$ of
kinetic equations \eqref{eq:kin} thus satisfies $J=\psi_T
\hat{\Gamma}_T(B) f - f_T\hat{\Gamma}_T(-B)\psi - j_S
(\psi_Tf_L+f_T\psi_L)=0$ and $\hat{\Gamma}_Lf=0$ at the NS interfaces
of the loop, for any $\psi$ that satisfies
$\hat{\mathcal{O}}(-B)\psi=0$, regardless of boundary conditions.

The only linear boundary condition consistent with the above is
$\hat{\Gamma}_Tf=G_T(\abs{B})f_T+j_Sf_L$, where $G_T$ describes
conductances related to the supercurrent conversion. For simplicity,
we then assume $G_T=\infty$ at $E<\abs{\Delta}$ and $G_T=0$ at
$E>\abs{\Delta}$, which results to
\begin{subequations}
\label{eq:vanishingqpcurrent}
\begin{align}
  \Gamma_L f &= 0 \,, & f_T &= 0 \,, && E<\abs{\Delta}\,, \\
  \Gamma_L f &= 0 \,, & \Gamma_T f &= j_S f_L \,, && E>\abs{\Delta}\,.
\end{align}
\end{subequations}
This acknowledges the fact that for $E<\abs{\Delta}$ the kinetic
equations imply a vanishing $f_T$ beyond the current conversion
region, and that in a BCS superconductor $f_T$ does not relax at
$E>\abs{\Delta}$ if there is no inelastic scattering.  \cite{schmid75}
Employing Eq.~\eqref{eq:vanishingqpcurrent} is analogous to requiring
that the ``non-equilibrium'' parts of the spectral currents vanish;
the remaining part $j_S f_L$ is what at equilibrium gives rise to the
supercurrent.

Note that Eq.~\eqref{eq:vanishingqpcurrent} is not exact: we at least
neglect the resistance in the supercurrent conversion region discussed
for example in Refs.~\cite{schmid75,boogaard2004-resistance}.  Note also
that when treating a superconducting loop as two boundary conditions,
charge conservation must be ensured by adjusting all potentials
relative to that of the superconductor. Nonetheless, we expect that
Eq.~\eqref{eq:vanishingqpcurrent} captures some of the relevant
physics in the problem.  Below, we use it to illustrate how charge
imbalance could change observable quantities.

\subsection{Conductance}
\label{subs:conductance}


How the proximity effect changes the conductance has been studied in
detail, both experimentally
\cite{pothier1994-flux-modulated,vegvar1994-mesoscopic,petrashov1995-phase,charlat96,courtois1996-long-range,hartog96,belzig2002-negative}
and theoretically
\cite{nazarov1996-diffusive,stoof1996-kinetic-equation,volkov96,golubov1997-coherent}.
For a review, see for example
Ref.~\cite{lambert1998-phase-coherent}.


The modification to conductance can conveniently be described with the
Usadel equations. Once $\tilde{L}_{\alpha\beta}^{ij}(E)$ is
known---usually the zeroth order in $j_S$ and $\T$ is accurate
enough---calculating various conductances can be done: one can
directly evaluate the corresponding conductance matrix $L_{11}^{ij}$
and thermoelectric coefficients $L_{12}^{ij}$ from
Eqs.~\eqref{eq:linearresponseL} and write
\begin{equation}
  \begin{split}
    \dd{I_c^i} =
    \sum_j L_{11}^{ij} \dd{V_j}
    + \sum_j L_{12}^{ij} \dd{T_j}/T
    \\
    + \sum_{j} \left.\frac{\partial I_c^i}{\partial \varphi_j}\right\rvert_{\{V\}=0,\{\varphi\}}\, \dd{\varphi_j}
  \,.
  \end{split}
\end{equation}
The second sum is finite if the heating of the terminals is
significant, but should still give only a small contribution as the
thermoelectric coupling is small, as can be seen in
Fig.~\ref{fig:Lmatrix}.  The last term arises if conductances are
evaluated in structures where the phases $\varphi_j$ in the
superconducting terminals may vary. However, for $i$ referring to a
normal terminal, $I_c^i(\{V\}=0,\{\varphi\})=0$ independent of the
phases $\{\varphi\}$. This implies that the last term vanishes for
conductances around $\{V\}=0$, the potential of the superconductors,
but it may be finite when calculating differential conductances.
Note also that when modeling superconducting loops using only
boundary conditions at the NS interfaces, current conservation needs
to be ensured by adjusting all potentials relative to that of the
superconducting condensate.

\begin{figure}\centering
  \includegraphics{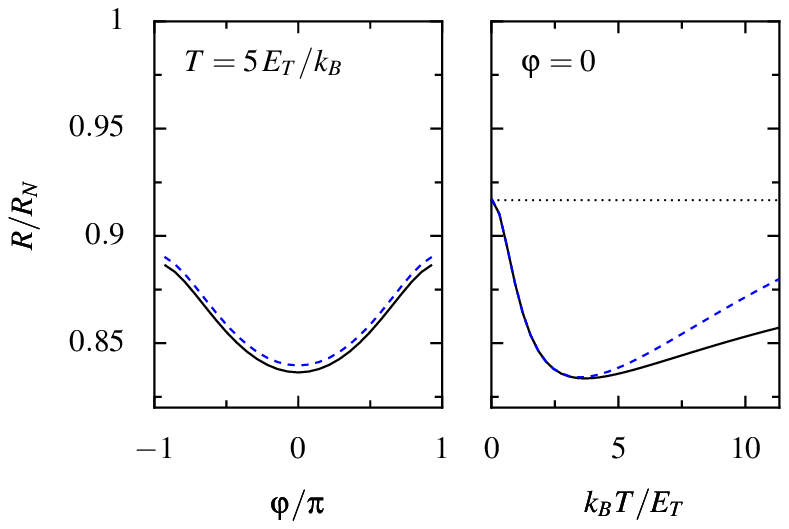}
  \caption{\label{fig:R} Linear--response electrical resistance $R$
    between terminals 1 and 2 of the structure in
    Fig.~\ref{fig:setup}(b), as a function of the phase difference
    $\varphi$ and the temperature $T$.  The resistances are normalized
    to the normal-state resistance $R_N = R_1 + R_2 + R_5$.
    The curves correspond to different models of the superconducting
    loop discussed in Sections~\ref{sec:interfaces}, \ref{subs:loops} --
    a) (solid) and b) (dashed). The current flows via the superconducting loop as
    supercurrent, reducing the resistance from the normal-state value
    also at $T=0$. Temperature dependence of the energy gap $\Delta$ is
    neglected, and we assume $\Delta=20E_T$.  }
\end{figure}

Typical behavior of conductance in an Andreev interferometer is
illustrated in Fig.~\ref{fig:R}. The proximity effect adds an
enhancement that oscillates with the superconducting phase difference
$\varphi$ and has a re-entrant dependence on the temperature $T$. The
figure also shows how charge transport via quasiparticles
($E>\abs{\Delta}$) in the superconducting loop may change the
conductance at high temperatures. The two curves correspond to the
terminal a) and long-loop b) limits discussed in
Sections~\ref{sec:interfaces}, \ref{subs:loops}.  For the former, the
loop contributes to electric conduction at energies $E>\abs{\Delta}$,
for the latter it does not.

\subsection{Thermal conductance}
\label{subs:thermalconductance}

\begin{figure}\centering
  \includegraphics{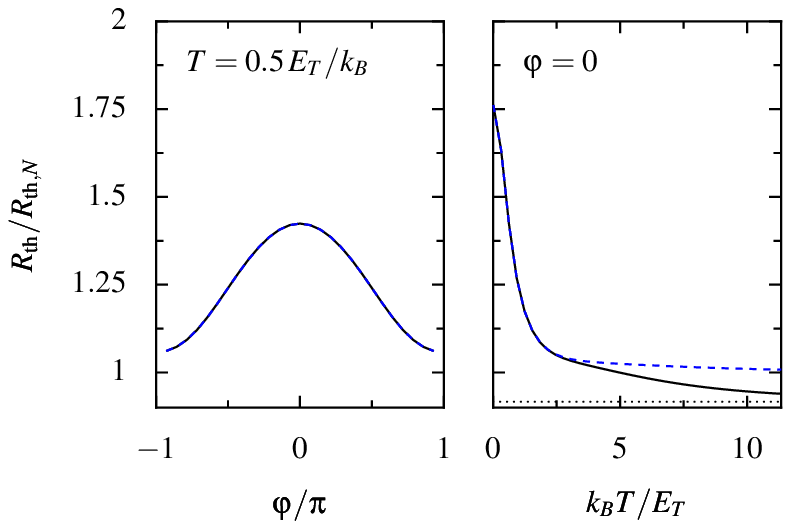}
  \caption{\label{fig:Rth} As Fig.~\ref{fig:R}, but the thermal
    resistance $R_{\mathrm{th}}$ is shown.  It is normalized to the
    normal-state Wiedemann--Franz value $R_{\mathrm{th},N} =
    3e^2 R_N/(\pi^2 k_B T)$. The two curves correspond to same models
    for the superconducting loop as in Fig.~\ref{fig:R}, a) terminal
    (solid) and b) long loop (dashed). In the former, at
    $k_B T\sim\Delta$, part of the thermal current flows through the loop
    as quasiparticle excitations, reducing the thermal resistance.
    Note that the scale for $R_{th}$ is the same in both figures.
  }
\end{figure}


As for the electrical conductance, the proximity of superconductors
modifies also the thermal conductance.
\cite{zhao2003-phase,bezuglyi03,jiang04b,jiang05} This was studied
on the basis of the quasiclassical Usadel theory in
Refs.~\cite{bezuglyi03,jiang04b}.

For a given setup, calculation of the thermal conductance from
$\tilde{L}_{\alpha\beta}^{ij}(E)$ proceeds as for the electrical
conductance.  Typical predicted features are $\varphi$-periodic
suppression of thermal conductance at low temperature $k_B
T\lesssim{}E_T$ due to modified density of states and thermal
diffusion coefficient $D\DL$, and inhibition of sub-gap thermal
transport into the superconductors due to Andreev reflection. These
are illustrated in Fig.~\ref{fig:Rth} for the example setup, together
with two models for the above-gap quasiparticle transport in the
superconducting loop.

\subsection{Thermopower}
\label{subs:thermopower}

\begin{figure}\centering
  \includegraphics{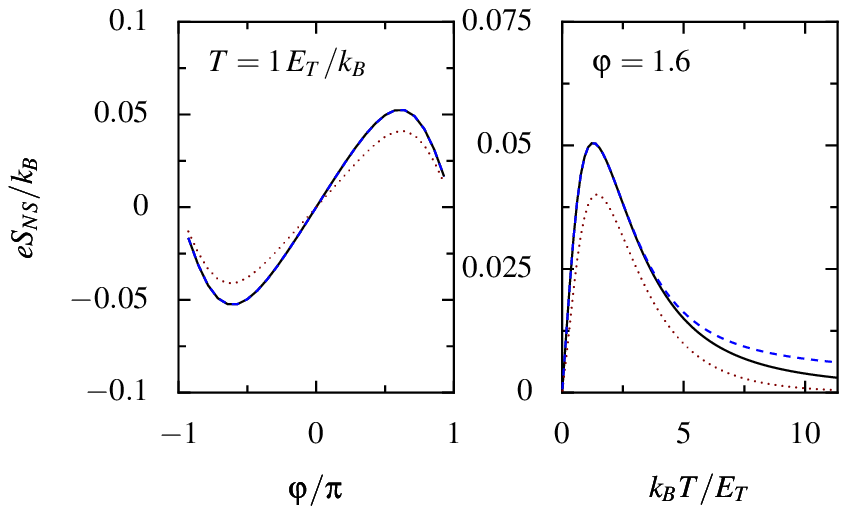}
  \caption{\label{fig:S} Linear-response thermopower in the structure
    of Fig.~\ref{fig:setup}(b), as a function of the phase--difference
    $\varphi$ and the temperature $T$.  Solid line: no charge
    imbalance in superconducting loop [case a) in
    Sec.~\ref{sec:interfaces}]. Dashed line: no inelastic relaxation
    in the long superconducting loop [case b) in
    Sec.~\ref{sec:interfaces}].  Dotted line: Approximation \eqref{eq:Sapprox},
    neglecting contributions from $\T$. If the terms proportional to
    $\T$ are taken into account, the result coincides with the solid
    line.  Other assumptions are as in Fig.~\ref{fig:R}.  }
\end{figure}

\begin{figure}\centering
  \includegraphics{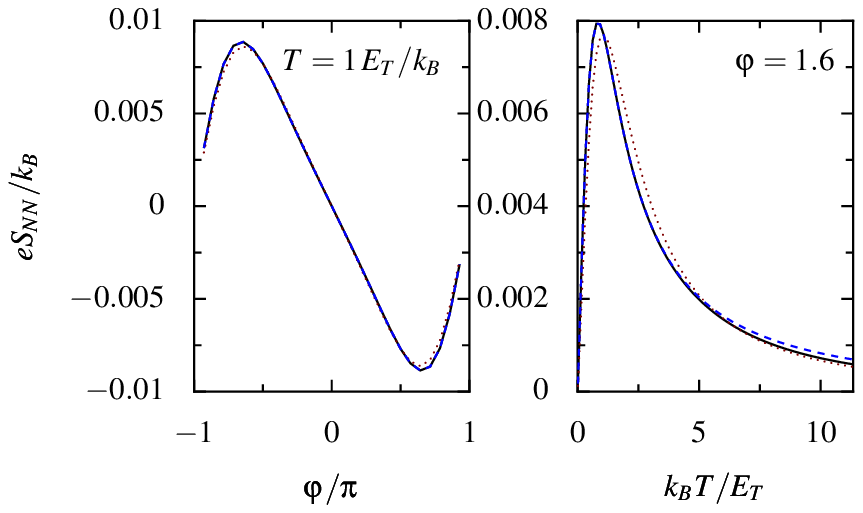}
  \caption{\label{fig:SNN} As Fig.~\ref{fig:S}, but showing the NN
    thermopower $S_{NN}$.  The dotted line includes the terms
    proportional to $\T$ in Eq.~\eqref{eq:Sapprox}; the other terms in
    Eq.~\eqref{eq:Sapprox} vanish.  }
\end{figure}



Thermopower $S$ is proportional to the upper right coefficient
$L_{12}$ of the thermoelectric matrix. The superconducting proximity
effect on $S$ has recently been studied experimentally, see
Refs.~\cite{dikin02,dikin02b,eom98,jiang05b,parsons03,parsons03b,zou2007-influence}
Theoretically, predictions for the thermopower in hybrid
normal--superconductor structures have been calculated starting from
the scattering theory in \cite{claughton96,heikkila00}, and via the
Usadel theory discussed here.
\cite{seviour00,kogan02,volkov05,virtanen04,virtanen04b} We discuss
the comparison between theory and the experiment in
Section~\ref{sec:discussion}, and consider here only the theoretical
model.


For a two-probe structure, the thermopower is usually defined as the
induced voltage divided by the temperature difference when no charge
current flows,
$S\equiv\left.\frac{\dd{V}}{\dd{T}}\right\rvert_{I_c=0}$, but the
additional terminals in the 4-probe structure in
Fig.~\ref{fig:setup}(b) allow for defining two distinct
thermopower-type quantities,
\begin{subequations}
\begin{align}
  S_{NS}&\equiv
  \left.\frac{\dd{(V_1 + V_2)}}{2\dd{(T_1 - T_2)}}\right\rvert_{I_{c,1}=I_{c,2}=0}
  \,,
  &
  S_{NN}&\equiv
  \left.\frac{\dd{(V_1 - V_2)}}{\dd{(T_1 - T_2)}}\right\rvert_{I_{c,1}=I_{c,2}=0}
  \,.
\end{align}
\end{subequations}
Both of these can be calculated from
$\tilde{L}_{\alpha\beta}^{ij}(E)$:
\begin{subequations}
\label{sec:SvsL}
\begin{align}
  S_{NS}
  &= T^{-1}\frac{1}{4} \mat{1 & 1} (L_{11}^{[12]})^{-1} L_{12}^{[12]}\mat{1\\-1}
  \,,\\
  S_{NN}
  &= T^{-1}\frac{1}{2} \mat{1 & -1} (L_{11}^{[12]})^{-1} L_{12}^{[12]}\mat{1\\-1}
  \,,
  \\
  L_{\alpha\beta}^{[12]} &\equiv \mat{
    L_{\alpha\beta}^{11} & L_{\alpha\beta}^{12} \\
    L_{\alpha\beta}^{21} & L_{\alpha\beta}^{22}
  }
  \,.
\end{align}
\end{subequations}
Typical results are shown in Figs.~\ref{fig:S} and~\ref{fig:SNN}. The
oscillations in $\varphi$ are always antisymmetric due to the symmetry
relation \eqref{eq:Lsymmetryqcl}, and the temperature dependence shows
the reentrant behavior on the energy scale of $E_T$ characteristic of
the superconducting proximity effect. One can also note that the
magnitude of the effect is significantly larger than what is expected
from the normal-state thermoelectric effects at sub-Kelvin
temperatures, which typically are of the order of
$S\approx10^{-4}$\ldots$10^{-3}\times{}k_B/e$.

Making use of expression \eqref{eq:linearization} and neglecting the
energy-dependence of $\DT$ and $\DL$ one can also derive
approximations such as \cite{virtanen04}
\begin{subequations}
\label{eq:Sapprox}
\begin{align}
  S_{NN}
  &\approx\frac{(R_3-R_4)R_5^2}{2(R_1+R_2+R_5)R_{SNS}}
  \frac{\dd{I_{S,eq}}}{\dd{T}}
  \\\notag
  &\quad
  + \frac{R_{SNS}(b_1+b_2)+(R_3+R_4)b_5}{(R_1+R_2+R_5)R_{SNS}}
  \,,
  \\
  S_{NS}
  &\approx
  \frac{4R_3R_4R_5 + R_5^2(R_3+R_4)}{4(R_1+R_2+R_5)R_{SNS}}
  \frac{\dd{I_{S,eq}}}{\dd{T}}
  \\\notag
  &\quad
  +\frac{R_{SNS}(b_1-b_2)+(R_3-R_4)b_5}{2(R_1+R_2+R_5)R_{SNS}}
  \,,
\end{align}
\end{subequations}
where $R_{SNS}=R_3+R_4+R_5$, $\abs{\Delta}\gg{}E_T$, and
\begin{align}
  b_j\equiv\int_0^\infty\frac{\dd{E}E}{2ek_BT^2}\sech^2\left(\frac{E}{2k_BT}\right)\frac{R_j}{L_j}\int_0^{L_j}\dd{x}\T(x)
\end{align}
are averages of the coefficient $\T$ in different wires. The
approximation \eqref{eq:Sapprox} is compared to the numerical solution
in Figs.~\ref{fig:S} and \ref{fig:SNN}. It turns out that a large part
of the thermopower is related to the equilibrium supercurrent
$I_{S,eq}$. \cite{seviour00,virtanen04} Note also that the
contribution from $I_S$ to $S_{NN}$ is strongly dependent on the
asymmetry in the structure and vanishes for a left-right symmetric
setup, as does the contribution from $\T$.
\cite{virtanen04,virtanen04b} However, the contribution from energies
$E>\abs{\Delta}$, which is neglected here, behaves differently in this
respect, see Refs.~\cite{kogan02,virtanen04b,volkov05}.

\subsection{Peltier effect}
\label{subs:peltier}

\begin{figure}\centering
  \includegraphics{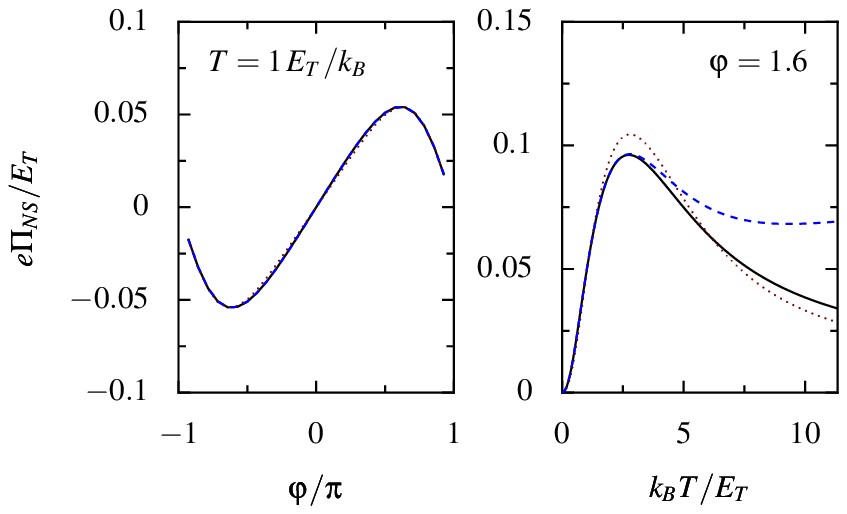}
  \caption{\label{fig:Pi}
    As Fig.~\ref{fig:S}, but showing the Peltier coefficient $\Pi_{NS}$.
    The dotted line is obtained from approximation \eqref{eq:Sapprox}
    including the $\T$-terms.
    The Kelvin relation $\Pi=TS$ can be seen by comparing to
    Fig.~\ref{fig:S}.
  }
\end{figure}

The second off-diagonal thermoelectric coefficient $L_{21}$ has not
yet been measured in the presence of the proximity effect, although
related experiments far from equilibrium have been made.
\cite{crosser06} Theoretical predictions for modifications due to the
proximity effect have been calculated from the scattering theory
\cite{claughton96} and from the Usadel theory \cite{virtanen07}.


A finite $L_{21}$ coefficient induces a Peltier effect, energy current
driven by charge current.  The Peltier coefficient $\Pi$ is in general
defined as the ratio of the heat current $I_Q=I_E-\mu I_c$ to the
charge current at constant temperature,
$\Pi\equiv\frac{\dd{I_Q}}{\dd{I_c}}$. In our example four-probe
structure in Fig.~\ref{fig:setup}(b), two Peltier coefficients can be
defined,
\begin{align}
  \Pi_{NS} &\equiv \left.\frac{\dd{I_E^1}}{\dd{I_c}}\right\rvert_{I_c^1=I_c^2=I_c/2}
  \,,
  &
  \Pi_{NN} &\equiv \left.\frac{\dd{I_E^1}}{\dd{I_c}}\right\rvert_{I_c^1=-I_c^2=I_c}
  \,,
\end{align}
corresponding to two different current configurations. These are directly
related to the linear-response $L$-coefficients by
\begin{subequations}
  \label{eq:pivsL}
\begin{align}
  \Pi_{NS}
  &=
  \frac{1}{4} \mat{1 & -1} L_{21}^{[12]} (L_{11}^{[12]})^{-1}
  \mat{1 \\ 1}
  \,,
  \\
  \Pi_{NN}
  &=
  \frac{1}{2} \mat{1 & -1} L_{21}^{[12]} (L_{11}^{[12]})^{-1}
  \mat{1 \\ -1}
  \,,
\end{align}
\end{subequations}
in a similar way as in Eq.~\eqref{sec:SvsL}.  However, note that
$\Pi_{NS}$ can be defined only when there is a grounded extra contact
in the superconducting loop [c.f. Fig.~\ref{fig:setup}(b)] through
which the injected current $I_c$ can flow.

As discussed above, the matrix element $L_{21}$ is usually coupled to
the element $L_{12}$ via Onsager's reciprocal relation. This leads to
Kelvin relations between the Peltier coefficients and the thermopower
\begin{align} \label{eq:kelvin}
  \Pi_{NS} &= T S_{NS} \,,
  &
  \Pi_{NN} &= T S_{NN} \,,
\end{align}
which are easily seen by transposing equations~\eqref{eq:pivsL} and
comparing to Eqs.~\eqref{sec:SvsL}.  These relations are not broken by
the superconducting proximity effect, which implies
that the proximity-induced Peltier coefficient inherits the magnitude,
phase oscillations and the temperature dependence of the thermopower.
Numerically calculated linear-response Peltier coefficient in the
example structure is illustrated in Fig.~\ref{fig:Pi}.

The Peltier coefficient is sufficiently large so that it could be
detected simply by observing how the effect changes the temperature
of one of the terminals in Fig.~\ref{fig:setup}(b). For a typical
Thouless energy $E_T/k_B=\unit[200]{mK}$, the coefficient in
Fig.~\ref{fig:Pi} achieves a magnitude of
$\Pi\sim\unit[1.5]{\mu{}V}$ at temperature $T=\unit[400]{mK}$. A
simple heat balance estimate, assuming that the terminal 1 is
thermally isolated apart from the electronic heat conduction through
wire 1,
\begin{align}
  I_Q^1 = -G_{th}\Delta T + 2 \Pi_{NS}I_c + e I_c^2/G = 0 \,,
\end{align}
then yields a maximum cooling $\Delta T \approx
-(3/\pi^2)(e^2\Pi_{NS}^2/k_B^2T)\sim\unit[0.2]{mK}$. However, the
oscillation amplitude is proportional to $I_c$ and can be larger than
this maximum cooling effect: variation of the order of millikelvin at
least should be possible. \cite{virtanen07} Temperature changes of
this order have already been successfully resolved in mesoscopic
structures \cite{meschke06}, so that in a suitably optimized
setup, it might also be possible to detect this proximity-Peltier
effect.

\section{Dependence on external flux} \label{sec:fluxdependence}
\authors{Tero and Pauli}


A magnetic field applied to a normal-metal--superconductor
heterostructure causes persistent currents to flow in the structure
and induces some dephasing. The currents also screen the applied
magnetic field, which can usually be taken into account by assigning
self-inductances to all loops in the structure.  Both effects can be
included in the present theory, and we discuss the latter briefly
below.

If considering the Andreev interferometer in
Fig.~\ref{fig:setup}(b), screening is mostly taken into account in
the $I_c(\varphi)$ relation of the weak link. The inductance $L$ of
the loop only modifies the $\varphi(\Phi_x)$ relation between the
induced phase difference $\varphi$ and the external magnetic field
$\Phi_x$ to \cite{tinkham,zou2007-influence}
\begin{align}\label{eq:phaseflux}
  \varphi - 2\pi\frac{\Phi_x}{\Phi_0} = L I_c(\varphi) \,.
\end{align}
One should note that although a modified $\varphi(\Phi_x)$ relation
should change the shape of the oscillation of various quantities as
functions of $\Phi_x$, e.g.  thermopower in Fig.~\ref{fig:S}, the
symmetry properties in Sec.~\ref{sec:symmetry} remain unchanged.
However, if there is hysteresis and multiple flux states are
possible for the same values of control parameters, the situation is
slightly more complicated: for a given solution of
\eqref{eq:phaseflux} with external flux $\Phi_x$, there exists a
solution with $-\Phi_x$ for which
Eqs.~(\ref{eq:onsagersymmetry},\ref{eq:Lsymmetryqcl}) apply.

There is a further effect of the magnetic field neglected in this
work: Zeeman effect, which leads effectively to an exchange field
inside the wires (for an example of such an effect, see
Ref.~\cite{heikkila00epl}). However, unless special care is taken,
this effect plays typically a much smaller role than the dephasing
effect of the field.

\section{Discussion}
\label{sec:discussion}
\authors{Pauli and Tero}

In this article, we have systematically discussed the predictions of
the quasiclassical diffusive-limit theory on the thermoelectric
response of normal-metal samples under the influence of the
proximity effect. The latter yields corrections to the fairly
general relations in Eq.~\eqref{eq:nstateexpressions}. These
corrections depend in general on energy (i.e., on temperature or
voltage) and on the phase difference between superconducting
contacts. At least in most typical cases, one of the general
relations, the Onsager relation (and thereby also the Kelvin
relation) holds also in the presence of the proximity effect.
Furthermore, the approximations made in the quasiclassical theory
imply that the diagonal coefficients of the thermoelectric matrix
are generally symmetric and the off-diagonal ones antisymmetric with
respect to an external magnetic flux.

Our results for the proximity correction of the conductance agrees
with the previous quasiclassical treatments
\cite{nazarov1996-diffusive,stoof1996-kinetic-equation,volkov96,golubov1997-coherent}.
However, as far as we know, the charge imbalance effect has not been
previously addressed. The thermal conductance calculated here is in
line with the results in Ref.~\cite{jiang05b}, but in contrast to
it, we do not make any approximations to the kinetic equations.

The quasiclassical prediction on the thermopower has been detailed
in different situations in
Refs.~\cite{seviour00,kogan02,volkov05,virtanen04,virtanen04b}. Our
theory is in line with these predictions. The mechanism for the
finite thermopower is analogous to the generation of charge
imbalance in bulk superconductors in the presence of coexisting
supercurrent and temperature gradient
\cite{pethick79,clarke79,schmid79,clarke80}.

To our knowledge, the only quasiclassical treatment of the Peltier
effect and the resulting temperature modification prior to this paper
is our Ref.~\cite{virtanen07}. Beyond the quasiclassical
approximation, these effects have been discussed using the scattering
theory and numerical simulations of the Bogoliubov -- de Gennes
equation on a tight-binding lattice \cite{claughton96,heikkila00}. In
that work, the symmetry of the flux dependence for the off-diagonal
coefficients was mostly dependent on the geometry and disorder of the
considered system, and not fixed as in our work. However, the small
size of the simulated structures makes a quantitative comparison for
example to the present work difficult: in Ref.~\cite{claughton96} even
the normal-state thermoelectric effects were large, and it is
difficult to distinguish those contributions from the proximity effect
that remain large in experimentally relevant structures from those
that rely on significant electron-hole asymmetry.

On the experimental side, a qualitative agreement to most of the
features presented here has been found. The resistance correction in
an Andreev interferometer has been found to oscillate with a
magnetic flux through the loop
\cite{pothier1994-flux-modulated,vegvar1994-mesoscopic,petrashov1995-phase,charlat96,courtois1996-long-range,hartog96,belzig2002-negative},
with the scale given by the flux quantum. Moreover, the reentrance
effect illustrated in Fig.~\ref{fig:R} has been measured in
different samples \cite{charlat96,hartog96,belzig2002-negative}.
However, to our knowledge there is no successful quantitative fit
between the quasiclassical predictions and the experimentally
measured temperature dependence of the resistance
--- see an example of such a comparison in
Ref.~\cite{belzig2002-negative}. The reason for this may be the
neglect of the generally temperature-dependent inelastic scattering
effects (see Sec.~\ref{subs:effectsleftout}) in the theory.

The thermopower in the presence of the proximity effect has been
measured by two groups, one in the Northwestern University, USA
\cite{dikin02,dikin02b,eom98,jiang05b}, and another in the Royal
Holloway University of London
\cite{parsons03,parsons03b,srivastava05,zou2007-influence}. Again,
most of the qualitative features agree with the quasiclassical
theory. The measured thermopower oscillates with the flux and is at
least two orders of magnitude larger than the normal-state
thermopower, and in line with the predictions from the
quasiclassical theory. The first attempt for a quantitative fit
\cite{zou2007-influence} of the temperature dependent thermopower
between the theory and the experiments was unsuccessful. We believe
that the major reasons for this were the too complicated geometry of
the measurements for this purpose and the neglect of the inelastic
scattering effects.

The major qualitative disagreement between the theory and the
measurement is in the symmetry of the thermopower oscillations with
the flux: in most measurements, the oscillations were antisymmetric
and in line with the theory
\cite{eom98,dikin02,dikin02b,jiang05b,parsons03,parsons03b,zou2007-influence},
in some measurements they were symmetric \cite{eom98,jiang05b}. The
authors of Ref.~\cite{jiang05b} suggested that this symmetry depends
on the geometry of the sample: in samples where the supercurrent
flows along with the temperature gradient, the oscillations are
antisymmetric whereas in other types of samples they are symmetric.
Such a conclusion cannot be made based on the quasiclassical theory.

We also note that in bulk superconductors, the magnitude of the
thermoelectric effects has been long under debate
\cite{ginzburgnobelrmp} --- there the experiments have shown larger
thermoelectric effects than those predicted by the theory.

The only published measurement on the thermal resistance $R_{\rm
th}$ of an Andreev interferometer known to us \cite{jiang05} showed
an oscillating $R_{\rm th}$, but the correction from the proximity
effect was larger than that predicted by the theory. We are not
aware of any measurements of the Peltier effect.

Quasiclassical theory, based on the combination of the BCS model and
the quasiclassical approximation, has been successful in providing a
quantitative explanation to a broad range of superconducting
phenomena. Here we have pointed out one qualitative aspect (flux
symmetry of the thermoelectric effects) which is yet to be
explained. Clearly, the full understanding of the nonequilibrium
electron transport phenomena in superconducting proximity samples
will still require both further experimental and theory work.

\appendix

\section{Possible reasons for the symmetric thermopower
oscillations}

In the diffusive limit, the antisymmetric flux dependence of the
proximity-induced thermopower results from the special symmetry of
the self-energies: all the typically relevant self-energies satisfy
Eq.~\eqref{eq:transformedselfenergy} in the presence of a magnetic
field $B$ with $\check{X}_2(B)=\check{X}_1(-B)$. Outside the
diffusive limit, one has to employ the Eilenberger equation
\cite{eilenberger68} describing the Keldysh Green's function
$\check{g}(\hat{p},\vec{r},E,B)$. Here $\hat{p}$ is the direction of
the electron momentum and $\vec{r}$ is the center-of-mass
coordinate. In this case, the property of the self-energies
$\check{x}[\check{g}]$ leading to the antisymmetric thermopower
oscillations is
\begin{equation}
\check{x}[\check{g}(\hat{p},\vec{r},E,B)]=-\hat{\tau}_1 \check{x}[-\hat{\tau}_1
\check{g}(-\hat{p},\vec{r},E,-B)\hat{\tau}_1]\hat{\tau}_1.
\end{equation}
This symmetry is satisfied for the most relevant self-energies,
including those for the elastic or spin-flip scattering in the Born
approximation, and that related to the superconducting order
parameter. We note that in Ref.~\cite{lofwander04}, it was shown
that a dilute concentration of impurities away from the Born limit
leads to large thermoelectric effects in unconventional
superconductors.

Beyond the quasiclassical approximation, other possible reasons for
the symmetric thermopower oscillations may be largely enhanced
electron-hole asymmetry effects (however, these were shown in
Refs.~\cite{wilhelm99,meyer99} to be small for a fairly generic
setup) or quantum interference contributions \cite{samuelsson01}.
Further studies on these effects are therefore required.

  This research was supported by the Finnish Cultural Foundation
  and the Academy of Finland.
  We thank N. Birge, M. Crosser, M.~Meschke and I.~A.~Sosnin for useful discussions.


\end{document}